\documentclass[onecolumn]{aa} 

\makeatletter
\renewcommand\@biblabel[1]{}
\makeatother
\usepackage{multirow}
\usepackage{graphicx}
\usepackage{epstopdf}
\usepackage{appendix}
\usepackage{txfonts}

\usepackage{setspace}
\usepackage{color}

\begin{document}

  \title{Direct measurement of desorption and diffusion energies of O and N atoms physisorbed on amorphous surfaces}
  \titlerunning{Desorption and diffusion energies of O and N atoms on amorphous surfaces}
  \subtitle{}

  \author{M. Minissale
         \inst{1}
         \and
         E. Congiu
         \inst{1}
         \and
         F. Dulieu
          \inst{1}
          }

  \institute{LERMA, Universit\'e de Cergy-Pontoise, Observatoire de Paris, ENS, UPMC, UMR 8112 du CNRS, 5 mail Gay Lussac, 95000 Cergy Pontoise Cedex, France\\
              \email{marco.minissale@obspm.fr}}

  \date{Received June, 2015, accepted November 2015}


 \abstract
  {Physisorbed atoms on the surface of interstellar dust grains play a central role in solid state astrochemistry. Their surface reactivity is one source of the observed molecular complexity in space. In experimental astrophysics, the high reactivity of atoms also constitutes an obstacle to measuring two of the fundamental properties in surface physics, namely desorption and diffusion energies, and so far direct measurements are non-existent for O and N atoms.}
  {We investigated the diffusion and desorption processes of O and N atoms on cold surfaces in order to give boundary conditions to astrochemical models.}
  {Here we propose a new technique for directly measuring the N- and O-atom mass signals. Including the experimental results in a simple model allows us to almost directly derive the desorption and diffusion barriers of N atoms on amorphous solid water ice (ASW) and O atoms on ASW and oxidized graphite.}
{We find a strong constraint on the values of desorption and thermal diffusion energy barriers. The measured barriers for O atoms are consistent with recent independent estimations and prove to be much higher than previously believed (E$_{des}=1410_{-160}^{+290}$; E$_{dif}=990_{-360}^{+530}$ K on ASW). As for oxygen atoms, we propose that the combination E$_{des}$-E$_{dif}$=1320-750 K is a sensible choice among the possible pairs of solutions. Also, we managed to measure the desorption and diffusion energy of N atoms for the first time (E$_{des}=720_{-80}^{+160}$; E$_{dif}=525_{-200}^{+260}$ K on ASW) in the thermal hopping regime and propose that the combination E$_{des}$-E$_{dif}$=720-400 K can be reasonably adopted in models. The value of E$_{dif}$ for N atoms is slightly lower than previously suggested, which implies that the N chemistry on dust grains might be richer.}
 {}


  \maketitle

%
\section{Introduction}

Molecules are often used as tracers of the physical conditions of many astrophysical environments. Their presence and abundance, as well as their internal states, give strong constraints on the temperature, density, and also dynamics of the observed media. Observations are almost always supported by models, and as soon as the abundance of a new molecular species can be measured, new or improved chemical pathways are implemented in these models. Gas phase chemistry can explain the majority of the observed molecular abundances, but there are cases where interstellar dust grains are indespensable to astrochemistry. This is the case for H$_2$ formation (Gould\&Salpeter 1963; Hollenbach\&Salpeter 1970), and for a long time, solid state chemistry has been thought to be the solution to the open questions gas phase chemistry could not address. For example, it has been proposed that hydrogenation, leading to saturation of species (i.e., C $\rightarrow$ CH$_4$, O $\rightarrow$ H$_2$O, CO $\rightarrow$ CH$_3$OH),  occurs on dust grains and subsequently is at the origin of part of the molecular complexity observed in space (Charnley et al. 2000).

Most astrochemical models that include solid-state chemistry (Hasegawa \& Herbst 1992; Charnley et al. 2000; Caselli et al. 2002; Cuppen \& Herbst 2007; Cazaux et al. 2010; Vasyunin et al. 2013; Reboussin et al. 2014 and references therein) refer to the article by Tielens \& Hagen (1982), a watershed paper that proposed diffusion controlled reactions and accretion/desorption processes as the corner-stone of solid state chemistry. 

\begin{figure}[t]
\centering
\includegraphics[width=9cm]{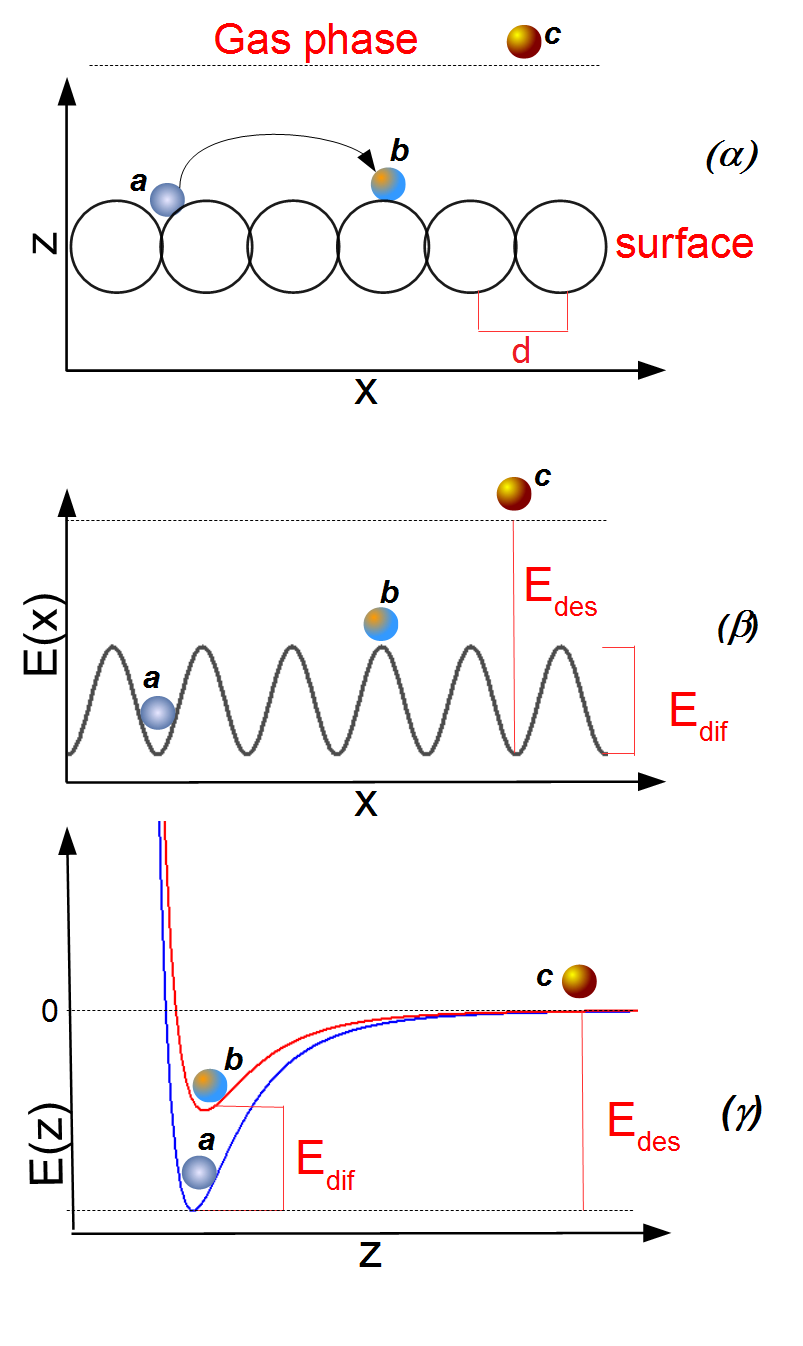}
\caption{\textit{\textbf{($\alpha$)}}: Schematic diagram of a substrate (black circles) with lattice constant d. \textit{\textbf{($\beta$)}}: potential energy of adparticles on a surface in the x direction. \textit{\textbf{($\gamma$)}}: potential energy of adparticles on a surface along the z direction. Particle (a) represents an adparticle in an adsorption site, particle (b) is in a transition state between two adsorption sites, and particle (c) represents a species in gas phase.} \label{fig:fig00}
\end{figure}

Atoms and molecules can adsorb on a surface through two kinds of bonding (chemisorption or physisorption) depending on the physical-chemical properties of the surface (temperature and chemical structure) and of the adparticle, i.e., its energy and electronic structure (Oura et al. 2003). 
At low surface temperatures where icy mantles can grow, physisorption processes involving Van der Waals interactions (10-400 meV) (Buckingham et al. 1988) are dominant. Physisorption concerns low surface temperatures ($<$ 100~K) and low particle energies ($<$ 50~meV), which are physical conditions that make physisorption hard to study experimentally and theoretically. Nevertheless in the last decade, an ever-growing number of works have dealt with a systematic study of the physics and chemistry of physisorbed species on cold non-metallic surfaces, namely graphite, silicates, and water ices (Pirronello et al. 1997; Ioppolo et al. 2008; Fillion et al. 2009; Goumans et al. 2010; Watanabe et al. 2010; Ward et al. 2011; Karssemeijer et al. 2012, 2014; Congiu et al. 2012; Thrower et al. 2014, and references therein). 
Both astrochemical models and experimental studies reveal that solid state chemistry is governed by a desorption-diffusion competition. 
In \figurename~\ref{fig:fig00} we show a sketch representing the desorption and diffusion processes and a diagram of the desorption energy E$_{des}$ and the diffusion energy E$_{dif}$ potentials.

If the desorption mechanism dominates, physisorbed reactive partners cannot increase the molecular complexity. Conversely, if diffusion mechanisms are preponderant, mobile atoms will be able to scan the surface and affect the abundance and variety of the species eventually created.
Diffusion and desorption barriers can be studied independently in the case of stable species (Mispelaer et al. 2013; Collings et al. 2004; Bisschop et al. 2006; Noble et al. 2012), but that is not the case for reactive species (i.e., atoms or radicals). In fact, the intrinsic mobility and reactivity of atoms governs the chemistry, but at the same time it hinders any direct measurement of the adsorption energy of atoms.

Experimental studies of reactive species were initially focused on the diffusion-desorption of H atoms (Manic\`o et al. 2001; Hornaeker et al. 2003; Amiaud et al. 2007). These works led to different results. The ambiguity probably originated in the use of an experimental technique (temperature-programmed desorption, TPD) that affects the mobility of atoms. For this reason, Matar et al. (2008) subsequently used the ability of H atoms to diffuse and react with a probe molecule (O$_2$) to study H mobility and obtain an effective diffusion barrier. In more sophisticated experiments, Hama et al. (2012) forced the desorption of H atoms by laser pulses and measured their residence time on different substrates to derive H mobility.

More recently, the diffusion of O atoms was investigated on different substrates experimentally and theoretically at low ($\sim$~6~K) (Minissale et al. 2013; Minissale et al. 2014a; Congiu et al. 2014; Lee\&Meuwly 2014) and high surface temperatures (50~K) (Minissale et al. 2015a). 
Simultaneously, two experimental groups (Ward et al. 2011; Kimber et al. 2014; He et al. 2015) proposed a rather high value for the adsorption energy of O atoms (i.e., 1500--1800~K) on different substrates, consistently with theoretical estimations by Bergeron et al. (2008). On the other hand, no experimental data have existed so far that cover N-atom desorption and diffusion barriers.

The aim of this article is twofold. First, we propose an original method of directly measuring the adsorption energy of reactive species; second, we provide an experimental value of the adsorption energies of O and N atoms on two surface analogues of astrophysical interest, i.e., compact amorphous solid water (ASW) ice and oxidized graphite. The ASW template mimics ice-coated grains in clouds with A$_v$>3, whereas oxidized graphite simulates grain surface conditions in thin clouds (A$_v$<3).


\section{Experimental}

Experiments were performed using the FORMOLISM set-up (Congiu et al. 2012), an ultra-high vacuum chamber (passivated for O atoms) coupled to a triply differentially pumped atomic beam aimed at a temperature controlled sample holder. The substrate is an ZYA-grade HOPG (highly oriented pyrolytic graphite) slab previously exposed to an O-atom
beam (oxidized) to avoid surface changes during the experimental sequences. 
Water ice films were grown on top of the HOPG substrate by spraying water vapour from a microchannel array doser located 2 cm in front of the surface. The water vapour was obtained from deionized water that had been purified by several freeze-pump-thaw cycles, carried out under vacuum. Amorphous solid water ice was dosed while the surface was held at a constant temperature of 110 K. 

O(N) atoms were produced by dissociating O$_2$(N$_2$) gas in a quartz tube placed within a Surfatron cavity. The cavity can deliver a maximum microwave power of 200~W at 2.45~GHz. We calibrated the atomic/molecular fluxes as described in Amiaud et al. (2007) and Noble et al. (2012). We found $\phi_O$=5$\pm$0.4$\times$10$^{12}$ atoms cm$^{-2}$s$^{-1}$, $\phi_N$=2$\pm$0.4$\times$10$^{12}$, and $\phi_{O_2, N_2}$=(3.0$\pm$0.3)$\times$10$^{12}$ molecules cm$^{-2}$s$^{-1}$. 
The dissociated fraction $\mu$ is typically 75\% for O$_2$ and 30\% for N$_2$. We can study the electronic state composition of the beam particles by tuning the energy of the ionizing electrons of the QMS. This technique allows ground-state or excited atoms and molecules to be selectively detected, as described in Congiu et al. (2009). We determined that our oxygen beam did not contain O($^1$D) nor O$_2$(a$^1\Delta^{-}_{g}$) and was composed of only O($^3$P) and O$_2$(X$^3 \Sigma^{-}_{g}$) (Minissale et al. 2014a). The nitrogen beam did not contain N atoms in an excited state, while a fraction of N$_2$ molecules were found to be in ro-vibrationally excited states (Minissale et al. 2014c). However, this is not an issue because atoms will relax rapidly on the surface. The beam temperature is typically lower than 400 K both for O and N atoms.

\subsection{The TP-DED technique}

\begin{figure}[t]
\centering
\includegraphics[width=9cm]{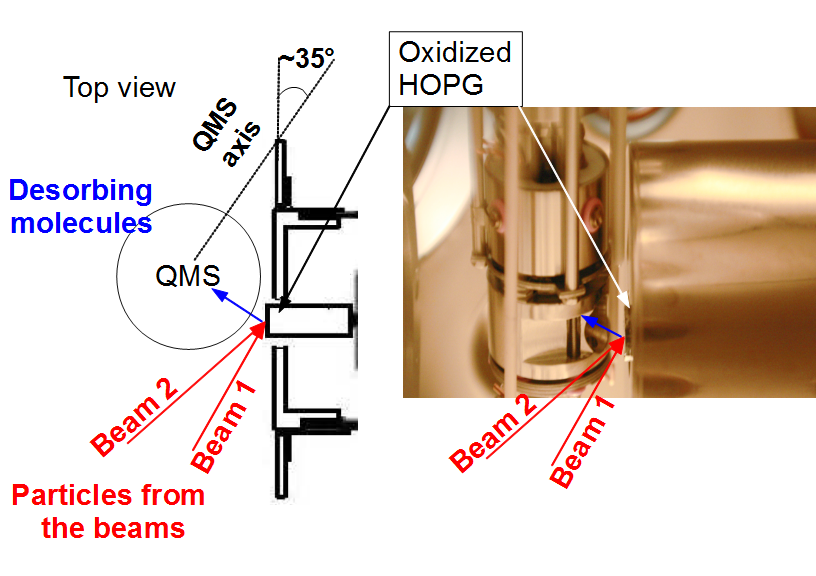}
\caption{Schematic top and side views (photo) showing the QMS and the cold surface configuration. The DED and TP-DED methods are indicated by red arrows (incoming particles from the beam lines) and by the blue arrow, which represents the fraction of desorbing species that are detected during the exposure phase. Adapted from Minissale (2014).} \label{fig:fig0}
\end{figure}

Experiments consisted in exposing the sample to the atomic beam while slowly reducing (17~mK/s) the sample temperature from 110~K to 10~K. The quadrupole mass spectrometer (QMS) was placed close to the surface and the species desorbing from the surface were measured continuously, as shown in \figurename~\ref{fig:fig0}.
Incoming atoms from the beam cannot enter the QMS head directly owing to the geometric configuration of our apparatus. Atoms can be scattered by the surface, desorb after thermalization, or react on the surface. The desorption flux is proportional to the mass signal measured in the experiments.
The new method we probed in this work is a combination of the King and Wells method, TPD technique, and during-exposure-desorption (DED) technique used to measure chemical desorption efficiency (Dulieu et al. 2013; Minissale et al. 2014b). The resulting combined technique is a temperature-programmed during-exposure desorption (TP-DED). The linearly decreasing temperature is an aspect of major importance. In fact, this allows us to have the surface free of adsorbates at least in the high-temperature part of the experiment. Unlike the coverage at high temperatures, the surface density is more questionable in the low-temperature regime. 

\subsection{Experimental results}

\begin{figure}[t]
\centering
\includegraphics[width=8.6cm]{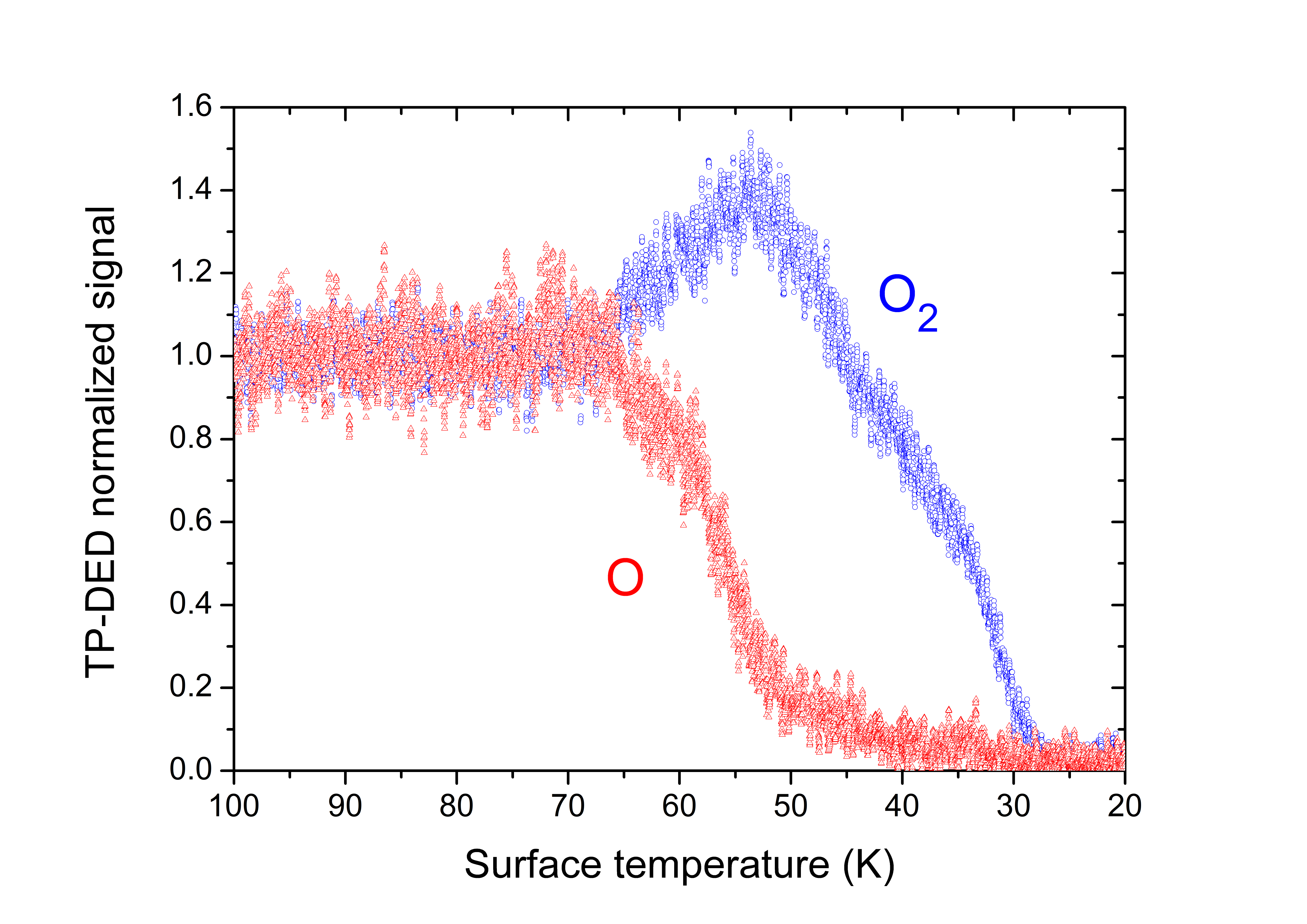}
\caption{TP-DED normalized signals at Mass 16 (red) and Mass 32 (blue) from compact ASW ice during exposure to the O beam. TP-DEDs were performed with a ramp of  -17~mK/s} \label{fig:fig1}
\end{figure}

An example of TP-DED is shown in \figurename~\ref{fig:fig1}. Here we show the normalized QMS ion count of mass 32 and mass 16 concurrent with an exposure of O atoms to compact ASW. Mass 32 and mass 16 correspond to O$_2$ and O detection, respectively. Because of cracking of molecules in the
QMS head, the contribution of O$_2$ to mass 16 is weak and falls within the error bars. Signals are normalized with respect to the high-temperature-regime ion count.
At high temperatures (above 68~K) we observe a plateau, which means that the number of atoms and molecules coming off the surface is constant. The plateau behaviour indicates that the desorption flux evens up with the incoming flux of particles. In other words, the residence time of O atoms on the surface is very short if compared with the mean time between the arrival of two atoms at the same adsorption site (about 300$\pm$25s, Minissale et al. 2014a). 

As long as the desorption rate is greater than the incoming flux, the accumulation of O atoms cannot occur. The desorption energy of O$_2$ molecules is considerably lower (Noble et al. 2012) than that of O atoms, therefore O$_2$ does not represent a high-coverage species at high temperatures. 
During the TP-DED, the desorption rate follows an exponential decrease with temperature. As soon as the accretion rate exceeds the desorption rate, the O-atom surface density increases. That compensates in part for the decrease in the desorption rate, but then diffusion-reaction events begin to dominate. 
This is why we observe the drop in atom count. The O atoms no longer leave the surface, but instead diffuse and react.
This is corroborated by the spectra of mass 32, showing an increase in the O$_2$ signal between 68 and 54~K. The residence time of O$_2$ at 68~K on compact ice is about 25 ms (Noble et al. 2012). Therefore, at our experimental timescales ($\gg$~1~s), O$_2$ desorbs almost immediately after it forms. Below 54~K, the signal at mass 32 begins to go down. The decrease in O$_2$ signal may be due to desorption, but is more likely related to its transformation into O$_3$, which is an efficient process as long as the O$_2$ residence time is long enough to allow for O+O$_2$ encounters (Minissale et al. 2014a). Actually, O and O$_2$ may be simultaneously on the surface for an interval of about 1~K, corresponding to a length of time of one minute under our experimental conditions. That would lead to the formation of at most 0.12 layers (upper limit) of ozone if we assume that the O+O$_2$ efficiency is one. 

At temperatures lower than 40~K, there is a second plateau for mass 16 that remains unchanged until the end of the cooling ramp occurring at 10~K. The intensity of this low-temperature plateau lies between 10 and 15$\%$ of the normalized signal. A simple interpretation of this residual O-atom signal is that it is due to the fraction of atoms bouncing off the surface without staying on it. This would indicate that the sticking coefficient of O atoms is higher than 85$\%$. This is consistent with what is expected. On one hand, sticking efficiencies of light species (H$_2$, HD, D$_2$) have shown to have a clear dependence on their mass (Matar et al. 2010; Chaabouni et al. 2012); on the other hand, heavier molecules like water or O$_2$ have a sticking efficiency higher than 90$\%$ (Bisschop et al. 2006). It must be noted that our spectrometer is located close to the specular reflection angle opposite the beam-line, thus the scattered part of the beam is probably overestimated, as we have seen in the case of O$_2$ molecules. Nonetheless, assuming a conservative lower limit of 85$\%$ for the sticking efficiency of O atoms seems reasonable.

In the case of nitrogen, the presence of N$_2$ in the lower plateau region is probably of more importance because the desorption energy values of N$_2$ and N are rather close. The total exposure time between the two plateaus, which is where most of the information is found (e.g., atoms and molecules present on the surface at once), corresponds to an integrated exposure of around two layers. 
Nevertheless, since the most molecules are desorbing at the beginning of the signal drop, we can fairly assume that no more than a complete layer was formed on the surface. This explains our choice of an unusual temperature ramp: slow enough to be certain that a quasi steady state was achieved at any time, but fast enough to be certain that no multilayer effects could occur.

\begin{figure*}[t]
\centering
\includegraphics[width=14cm]{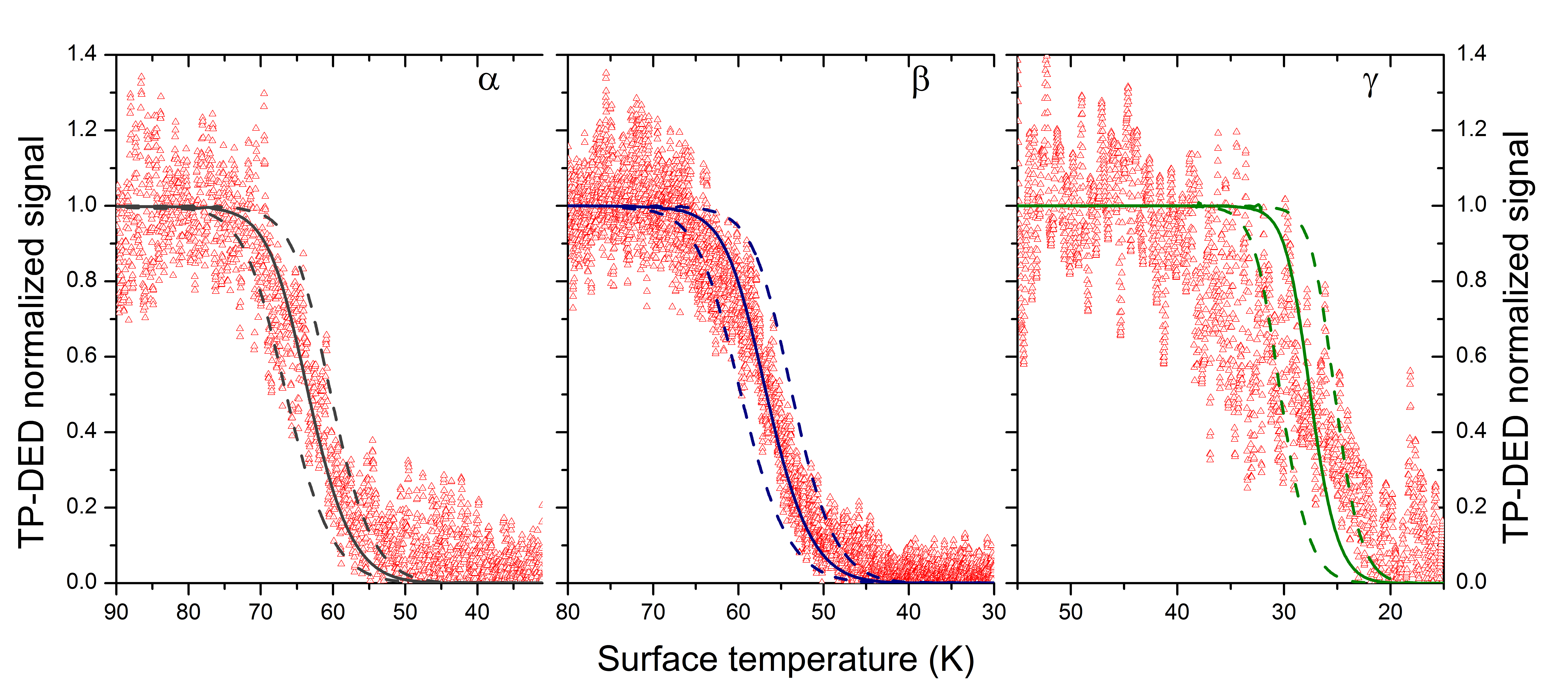}
\caption{TP-DED normalized signal as a function of surface temperature (thermal ramp = $- 17$ mK/s) of O atoms exposed to oxidized graphite (panel $\alpha$), O (panel $\beta$) and N atoms (panel $\gamma$) exposed to compact ASW. The curves represent modeling results obtained with different combinations of E$_{des}$--E$_{dif}$ values, each treated as a free parameter.} \label{fig:fig2}
\end{figure*}

\figurename~\ref{fig:fig2} displays the experimental and modeled TP-DED spectra obtained during irradiation with O atoms of oxidized graphite (panel $\alpha$) and irradiation with O atoms and N atoms of ASW (panel $\beta$ and panel $\gamma$, respectively). In each case, the trend of the experimental data follow the same scheme:
\begin{itemize}
        \item  a high-temperature plateau, i.e., desorption flux $=$ incoming beam of particles;
        \item  an intermediate region, where diffusion-reaction events consume atoms and the desorption flux is reduced with decreasing temperature;
        \item  a low-temperature plateau, where atoms no longer desorb from the surface (except for a small fraction bouncing off the surface without accommodating on it).
\end{itemize}

The differences in the three cases can be interpreted according to the position of high- and low-temperature plateaus. Experimental values of O atoms sent onto ASW show that the spectral features are shifted towards lower temperatures with respect to the results obtained on oxidized graphite. If we compare the curves of N and O obtained on ASW, we notice that the N plateaus come at lower temperatures. 

\section{Discussion}

It is possible to build a very simple model to reproduce the experimental curves. This was done by computing the evolution of the atom population on the surface. The loss rate of species X from the surface is
\begin{equation}
\frac{dX}{dt}= \phi_X - X k_{X-des} - R(X,k_{X-dif})
\end{equation}
where $\phi_X$ is the flux of incoming atoms, $X$ is the surface density of atom X, and
$k_{X-des}$ is the desorption rate  
\begin{equation}\label{eq:des}
k_{X-des}=\nu_0 e^{-E_{X-des}/k_bT_s}
\end{equation} 
where $\nu_0$ is the vibration frequency typically assumed to be 10$^{12}$~s$^{-1}$, $T_s$ is the surface temperature, and E$_{X-des}$ represents the desorption energy. The resulting desorption flux  $X\cdot k_{X-des}$ is proportional to the mass signal measured in the experiments. 
Finally, R(X,k$_{X-dif}$) represents a series of terms accounting for reactions involving particle X (Minissale et al. 2014a, 2015a), where X can be O, O$_2$, or O$_3$, and k$_{X-dif}$ and E$_{X-dif}$ are the diffusion rate and diffusion energy of species X, respectively.
Since diffusion should be dominated by thermal hopping in the range of high temperatures (30--70~K), we use an Arrhenius-type law (eq. \ref{eq:des}) to simulate diffusion by using E$_{X-dif}$ instead of E$_{X-des}$. E$_{X-des}$ and E$_{X-dif}$ represent the two numerical parameters of our model; the symbol X will be omitted when we do not refer to a specific species or whenever that is clearly specified in the text. As for O$_2$, O$_3$, and N$_2$, the diffusion and desorption energies were fixed. Their values are shown in Table 1.
\begin{table}[ht]
\centering
\caption{Diffusion and desorption energies on oxidized graphite and ASW ice for O$_2$, O$_3$, and N$_2$ used in the model. The E$_{dif}$/E$_{des}$ ratio was fixed to 0.7}\label{tab:2}
\begin{tabular}{l | c | c}
\hline\hline
Species & Ox-Graph & ASW\\
        & \multicolumn{2}{c}{ E$_{dif}$--E$_{des}$ (K)}\\
\hline
O$_2$& 890--1260$^{a,b}$ & 820--1160$^a$\\
O$_3$& 1470--2100$^b$   & 1260--1800$^c$\\
N$_2$&  --              & 810--1150$^b$\\
\hline\hline
\end{tabular}
\scriptsize \\$^a$ Noble et al. 2012, Minissale 2014, Cuppen\&Herbst 2007.
\end{table}

\begin{figure}[t]
\centering
\includegraphics[width=9cm]{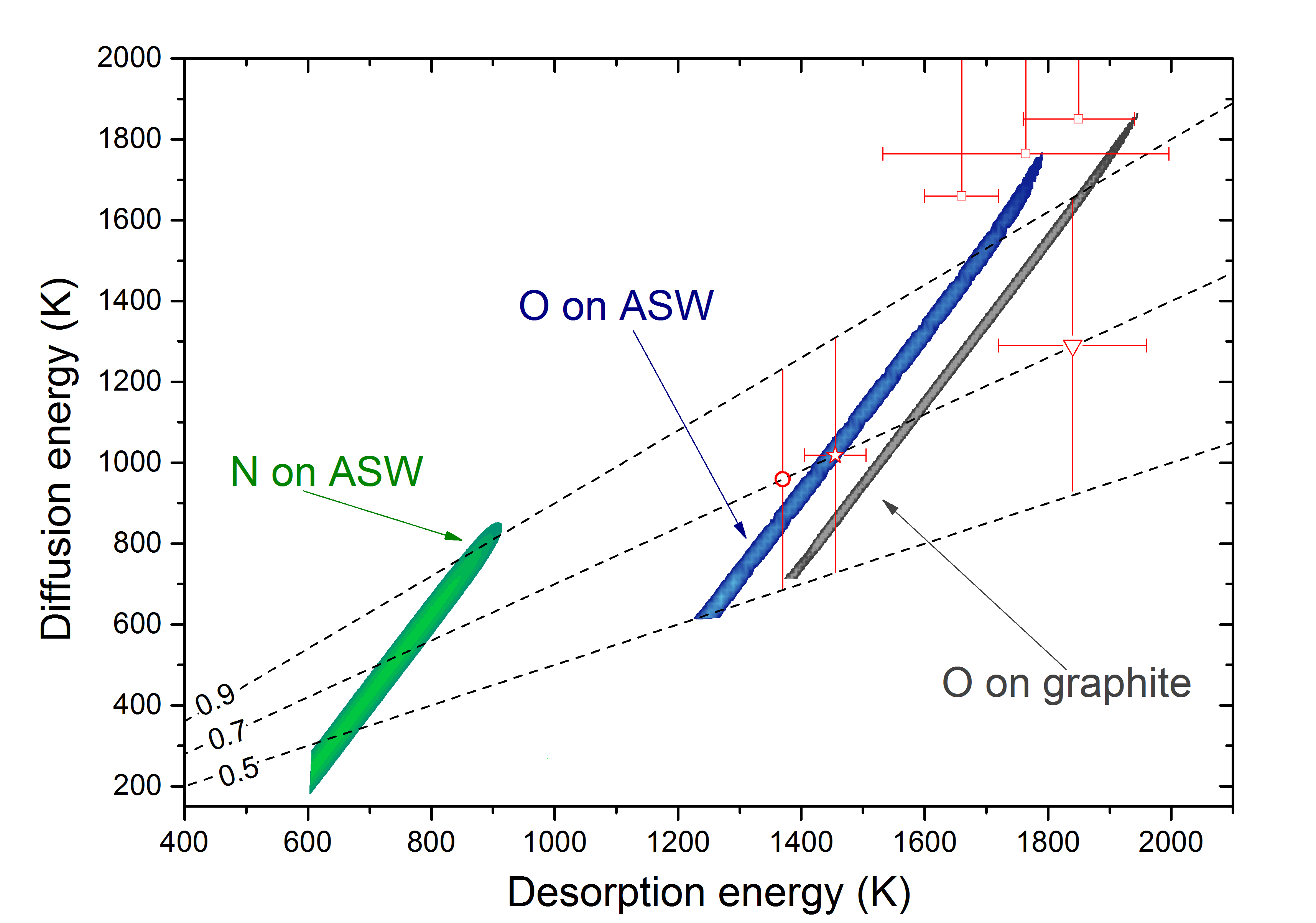}
\caption{Diffusion energies {\it vs.} desorption energies obtained for O and N atoms (blue and green bands, respectively) on ASW ice, and O atoms (grey band) obtained on oxidized graphite. Dotted lines represent E$_{dif}$/E$_{des}$~=~0.5, 0.7, and 0.9. 
Open symbols represent previous estimations of E$_{des}$ for O atoms: 1350~K calculated by Bergeron et al. (2008) on pyrene; 1455~K found Ward\&Price (2011) on C$_2$H$_2$-coated graphite; 1840~K measured by Kimber et al. (2014) on propyne ice; 1764~K found by He et al. (2014) on amorphous silicate; 1660~K and 1850~K measured by He et al. (2015) on porous water ice and amorphous silicate, respectively.} \label{fig:fig3}
\end{figure}

The curves in \figurename~\ref{fig:fig2} represent our modeling results. We found that different couples of E$_{des}$ and E$_{dif}$  exist that lead to a good fit of the data. If E$_{des}$ is high, the plateaus tend to appear at high temperature, while if E$_{des}$ value is reduced, then the plateaus shift towards lower temperatures. Also E$_{dif}$ affects the position of the plateaus, but this effect is smaller. Increasing E$_{dif}$ also shifts the plateaus towards high temperatures. 
It is therefore possible to get a reasonable fit for several combinations of E$_{des}$ and E$_{dif}$. 
\figurename~\ref{fig:fig3} shows the possible set of values of the E$_{des}$--E$_{dif}$ pairs that are able to reproduce our experimental values. In the E$_{des}$ {\it vs.} E$_{dif}$ graph, the coloured bands represent the regions where the 
$\chi^2$ lies within 10$\%$ of its minimum. We discarded the values of diffusion energy lower than 500~K for O atoms, since thermal diffusion should not be below this value at low temperatures (Congiu et al. 2014). On the other hand, in the upper region of the bands, high values are systematically eliminated by the filter that discards fits of poor quality.

The bands indicating the acceptable couples of solutions are fairly elongated and almost linear. The desorption barrier is relatively more constrained than the diffusion barrier. The slope of the bands is close to 2 for O atoms (blue and grey bands): for an increase in 1~K of desorption energy, the diffusion energy rises by 2~K. Overall, E$_{dif}$/E$_{des}$ ratios are between 0.5 and 0.9, as shown by the dotted lines in \figurename~\ref{fig:fig3}. 
This is the first experimental evidence that the value E$_{dif}$/E$_{des} \sim 0.5$, which is usually adopted in models of astrochemistry (Caselli et al. 2002; Cazaux \& Tielens 2004; Cuppen \& Herbst 2007), is a plausible choice. Our experiments also rule out the possibility of having values of E$_{dif}$/E$_{des}$ below 0.5 and above $0.9$.

Table 2 gives some examples of diffusion-desorption energy combinations corresponding to E$_{dif}$/E$_{des}$ ratios of 0.5, 0.7, and 0.9 obtained for O and N on the two substrates investigated in this work. 
One can adopt --- according to an educated guess or a measured value of either E$_{des}$ or E$_{dif}$ --- any of the couples satisfying the fits of \figurename~\ref{fig:fig3}. These are not to be considered true error bars, because once a value of either E$_{des}$ or E$_{dif}$ has been chosen, the other is well constrained.

\begin{table}[ht]
\centering
\caption{Diffusion and desorption barriers obtained for O and N atoms corresponding to E$_{dif}$/E$_{des}$ ratios =~0.5, 0.7, 0.9.}\label{tab:1}
\begin{tabular}{l|c|c|c}
  \hline\hline
 E$_{dif}$/E$_{des}$        & 0.5 & 0.7 & 0.9\\
                  \hline
                &\multicolumn{3}{c}{ E$_{dif}$--E$_{des}$ (K)}   \\
  \hline
  O on Ox-Graph & $690-1380$ & $1100-1580$ & $1675-1860$ \\
  O on ASW      & $625-1250$ & $990-1410$  & $1520-1700$  \\
  N on ASW      & $320-640$ &  $525-720$  & $790-880$  \\
  \hline\hline
\end{tabular}
\end{table}
An O diffusion barrier in the range $690 - 1100$~K (0.5$<$E$_{dif}$/E$_{des}$$<$0.7) is perfectly consistent with what we found previously (values between 600 and 900~K) to reproduce the efficiency of the H$_2$CO+O reaction (Minissale et al. 2015a). Also, desorption energies for O are consistent with previous estimations (Bergeron et al. 2008; Ward et al. 2011; Kimber et al. 2014; He et al. 2015); He et al. (2015) found 1850 $\pm$~90 K on amorphous silicates and 1660 $\pm$~60 K on porous water ice (see open-symbol points in \figurename~\ref{fig:fig3}).
The values of E$_{des}$ we derived for O atoms on oxidized graphite cover the interval of values (1455$\pm$72~K) proposed by Ward et al. (2011) and Kimber et al. (2014) obtained on an analogous sample, coated with C$_2$H$_2$. Also the theoretical value of 1350~K obtained for physisorption of O on pyrene by Bergeron et al. (2008) is compatible with our results. Since Bergeron et al. (2008), Ward et al. (2011), and Kimber et al. (2014) do not suggest an explicit value for the diffusion energy, the corresponding E$_{dif}$ values displayed in \figurename~\ref{fig:fig3} were drawn according to a E$_{dif}$/E$_{des}$ ratio of 0.7.
The good agreement of our fits with estimations of other authors, though achieved on different substrates, proves the reliability of the new direct method we present. Moreover, we provide the first evaluation of E$_{des}$ and E$_{dif}$ for N atoms. They are lower than the ones found for O atoms. In fact, the same difference existing between E$_{des}$ and E$_{dif}$ of N and O atoms appears if one considers desorption and diffusion barriers of the molecular species N$_2$ and O$_2$. Oxygen atoms are generally more bound to the substrate than O$_2$ molecules, indicating that the polarizability is not the only parameter that governs the adsorption energy. The high electronegativity of O atoms could perhaps explain that behavior. Another explanation could be that O($^3$P) atoms have unpaired electrons that generate a quadrupole moment. In contrast, molecular and atomic nitrogen seem to have adsorption energies in the same energy range.

Both substrates used in this work are amorphous and possess a complex distribution of adsorption and diffusion barriers. However, the technique we used implies that the experiments are carried out under conditions of very low surface coverage, where diffusion leads the atoms to occupy the minima on the surface potential. Therefore, desorption energies derived here are very likely to be related to the deepest sites, which means that we probed the high-energy end of the barrier distribution. That our modeling results fit the data is a clear indication that choosing unique values for desorption and diffusion is a reasonable choice in our case, notwithstanding the existence of an energy distribution.

In astrochemical models, the diffusion of species is very often directly and simply assumed to be linked to the desorption energy. Because all the chemistry is ruled by diffusion/reaction and diffusion/desorption competitions, this strong hypothesis has deep consequences. Such an assumption is also due to the poor knowledge of diffusion of molecules on a complex surface. Recently, Karssemeijer\&Cuppen (2014) have shown calculations on the diffusion of CO and CO$_2$ on water ice surfaces. They find that the E$_{dif}$/E$_{des}$ ratio lies within the 0.3-0.4 range, rather than between 0.5 and 0.9 as we found in this work. We cannot directly compare their calculations with our work because, firstly, the two studies investigate different systems and, secondly, atoms and molecules are likely to have different diffusion properties. Moreover, as we mentioned earlier, we probably probe only the high-value tail of the distribution of binding energies and also the high end of the diffusion barrier distribution if the two energies are directly coupled. 
However, we can limit the range for E$_{dif}$/E$_{des}$ ratio by using indirect constraints on O-atom diffusion, derived from our previous study of the H$_2$CO+O system (Minissale et al. 2015a). In this case, experiments were performed on a substrate held at 55 K corresponding to a thermal-hopping-dominated regime. The substrate was ASW ice or oxidized graphite, coated with H$_2$CO ice. In the H$_2$CO+O system, the key point is not the desorption vs diffusion competition, but diffusion vs reactivity. We succeeded in reproducing our data using an O-diffusion barrier ranging from 600~K to 900~K. If we choose the mean value (750 K) of this interval, and use it to constrain the results of the present paper, we have a E$_{dif}$-E$_{des}$ =~750-1320~K for ASW ice and 750-1420~K for oxidized graphite. We notice that the two E$_{dif}$/E$_{des}$ ratios are 0.57 and 0.53, respectively. Therefore, in the absence of other experimental evidence, we conclude that the most likely number to assign to the E$_{dif}$/E$_{des}$ ratio is 0.55. By using this value, in the case of N atoms, we find E$_{dif}$-E$_{des}$ =~400-720~K.
We realize that the diffusion/desorption energy ratio of 0.55 we derive does not come at the end of an explicit demonstration, but it is a conclusion we draw after the analysis of a series of converging facts. In the future more evidence will be available we believe, for example thanks to extended calculations, and our proposed value will be put to the test to verify whether it is plausible or not as far as atoms are concerned. It must be noted, however, that the E$_{dif}$/E$_{des}$ =~0.55 we propose represents the reference value of current astrochemical models. Our findings show that the assumptions made in the present codes are meaningful and reasonable, at least for the cases of O and N atoms.


\section{Conclusions}

In this work, we present a new method (TP-DED)of deriving desorption and diffusion energies of reactive species. We validated our method by modeling the experimental values and comparing our results to previous calculations and measurements of O-atom desorption energies. For the first time, we provide an estimation of N-atom diffusion and desorption energies on ASW ice. These findings are of major interest to astrophysicists because astrochemical models simulating solid state physics-chemistry on dust grain surfaces will implement the new values of diffusion-desorption energies of O and N atoms in their codes and perhaps improve our understanding of the observed interstellar abundances. 

We discussed the empiric relation between diffusion and desorption barriers and confirmed experimentally that 0.5 is a reasonable choice for diffusion-desorption barrier ratios of atoms, showing that the initial guess by Tielens \& Hagen (1982) was prophetic. We have proposed a value of 0.55, although there may be cases where E$_{dif}$/E$_{des}$ ratios ranging between 0.5 and 0.9 are a more realistic choice for atoms. As for molecules, the 0.3 -~0.4 range derived from calculations by Karssemeijer \& Cuppen (2014) has no reason to be questioned, and we would retain it among the possible E$_{dif}$/E$_{des}$ values characterizing molecular species. Here we proposed the combination E$_{dif}$-E$_{des}$ =~400-720 K for N atoms on icy mantles.
The 400~K diffusion energy of nitrogen atoms may have important implications for interstellar chemistry. In fact, it is low enough to allow for fast mobility at temperatures around 15~K, which is a diffusion rate of more than one hop per second. N atoms thus might be able to scan a large section of the grain surface before any other accretion event occurs. On the other hand, the 720~K desorption energy value is high enough to allow for a long residence time at the same interval of temperatures. While the N+N reaction is unlikely on dust grains, the most probable reaction involving N atoms is N+H $\rightarrow$ NH --- as was shown experimentally by Fedoseev et al. (2015) --- or N+H$_2$ (which should have an activation barrier). As a final speculation, we would like to emphasize that the synthesis of NH on bare grain surfaces should lead to a high percentage (about 45\%) of NH radicals released in the gas phase upon formation through the chemical desorption process (Minissale et al 2015b).

\begin{acknowledgements}
We acknowledge the support of the national PCMI programme founded by the CNRS, and DIM ACAV of the Conseil Regional d'Ile de France. MM acknowledges LASSIE Seventh Framework Programme ITN under Grant Agreement No. 238258. We thank our colleagues at the LERMA and ISMO laboratories for fruitful discussions.
\end{acknowledgements}

\end{document}